\documentclass[longauth,letter,traditabstract]{aa} 
\usepackage{graphicx}
\usepackage{txfonts}
\usepackage{upgreek}
\usepackage{tipa}
\usepackage{verbatim}
\usepackage{natbib}
\usepackage{color}
\bibpunct{(}{)}{;}{a}{}{,}

\newcommand{\source}{W3~IRS5}
\newcommand{\hoA}{o-H$_2^{17}$O 1$_{10}$-1$_{01}$}
\newcommand{\hoB}{p-H$_2$O 2$_{02}$-1$_{11}$}
\newcommand{\hoC}{p-H$_2^{18}$O 1$_{11}$-0$_{00}$}
\newcommand{\hoD}{p-H$_2$O 1$_{11}$-0$_{00}$}
\newcommand{\hoE}{o-H$_2$O 2$_{21}$-2$_{12}$}
\newcommand{\hoF}{o-H$_2$O 2$_{12}$-1$_{01}$}
\newcommand{\kms}{km~s$^{-1}$}

\newcommand{\msun}{$\rm M_{\odot}$}

\newcommand{\be}{\begin{equation}}
\newcommand{\ee}{\end{equation}}
\newcommand{\bd}{\begin{displaymath}}
\newcommand{\ed}{\end{displaymath}}
\newcommand{\bi}{\begin{itemize}}
\newcommand{\ei}{\end{itemize}}
\newcommand{\bfig}{\begin{figure}}
\newcommand{\efig}{\end{figure}}
\newcommand{\bc}{\begin{center}}
\newcommand{\ec}{\end{center}}

\newcommand{\vlsr}{$V_{\mathrm{LSR}}$}
\newcommand{\vtur}{$V_{\textrm{\tiny{tur}}}$}
\newcommand{\vexp}{$V_{\textrm{\tiny{exp}}}$}

\newcommand{\lsun}{$L_{\odot}$}

\newcommand{\agua}{$X_{\textrm{\tiny{H$_2$O}}}$}

\begin{document}
\title{Water in massive star-forming regions: \\HIFI observations of \source \thanks{Herschel is an ESA space observatory with science instruments provided by European-led Principal Investigator consortia and with important participation from NASA.}}

\author{
L.~Chavarr\'{\i}a\inst{\ref{inst1}} 
\and F.~Herpin\inst{\ref{inst1}}
\and T.~Jacq\inst{\ref{inst1}}
\and J.~Braine\inst{\ref{inst1}}
\and S.~Bontemps\inst{\ref{inst1}}
\and A.~Baudry\inst{\ref{inst1}}
\and M.~Marseille\inst{\ref{inst10}}
\and F.~van der Tak\inst{\ref{inst10},\ref{inst11}}
\and B.~Pietropaoli\inst{\ref{inst1},\ref{inst46}} 
\and F.~Wyrowski\inst{\ref{inst30}}
\and R.~Shipman\inst{\ref{inst10}}
\and W.~Frieswijk\inst{\ref{inst10}}
\and E.F.~van~Dishoeck\inst{\ref{inst6},\ref{inst2}}
\and J.~Cernicharo\inst{\ref{inst16}}
\and R.~Bachiller\inst{\ref{inst12}}
\and M.~Benedettini\inst{\ref{inst13}}
\and A.O.~Benz\inst{\ref{inst3}}
\and E.~Bergin\inst{\ref{inst14}}
\and P.~Bjerkeli\inst{\ref{inst9}}
\and G.A.~Blake\inst{\ref{inst15}}
\and S.~Bruderer\inst{\ref{inst3}}
\and P.~Caselli\inst{\ref{inst4},\ref{inst5}}
\and C.~Codella\inst{\ref{inst5}}
\and F.~Daniel\inst{\ref{inst16}}
\and A.M.~di~Giorgio\inst{\ref{inst13}}
\and C.~Dominik\inst{\ref{inst17},\ref{inst18}}
\and S.D.~Doty\inst{\ref{inst19}}
\and P.~Encrenaz\inst{\ref{inst20}}
\and M.~Fich\inst{\ref{inst21}}
\and A.~Fuente\inst{\ref{inst22}}
\and T.~Giannini\inst{\ref{inst23}}
\and J.R.~Goicoechea\inst{\ref{inst16}}
\and Th.~de~Graauw\inst{\ref{inst10}}
\and P.~Hartogh\inst{\ref{inst49}}
\and F.~Helmich\inst{\ref{inst10}}
\and G.J.~Herczeg\inst{\ref{inst2}}
\and M.R.~Hogerheijde\inst{\ref{inst6}}
\and D.~Johnstone\inst{\ref{inst7},\ref{inst8}}
\and J.K.~J{\o}rgensen\inst{\ref{inst24}}
\and L.E.~Kristensen\inst{\ref{inst6}}
\and B.~Larsson\inst{\ref{inst25}}
\and D.~Lis\inst{\ref{inst26}}
\and R.~Liseau\inst{\ref{inst9}}
\and C.~M$^{\textrm c}$Coey\inst{\ref{inst21},\ref{inst27}}
\and G.~Melnick\inst{\ref{inst28}}
\and B.~Nisini\inst{\ref{inst23}}
\and M.~Olberg\inst{\ref{inst9}}
\and B.~Parise\inst{\ref{inst30}}
\and J.C.~Pearson\inst{\ref{inst31}}
\and R.~Plume\inst{\ref{inst32}}
\and C.~Risacher\inst{\ref{inst10}}
\and J.~Santiago-Garc\'{i}a\inst{\ref{inst33}}
\and P.~Saraceno\inst{\ref{inst13}}
\and J.~Stutzki\inst{\ref{inst47}}
\and R.~Szczerba\inst{\ref{inst48}}
\and M.~Tafalla\inst{\ref{inst12}}
\and A.~Tielens\inst{\ref{inst6}}
\and T.A.~van~Kempen\inst{\ref{inst28}}
\and R.~Visser\inst{\ref{inst6}}
\and S.F.~Wampfler\inst{\ref{inst3}}
\and J.~Willem\inst{\ref{inst10}}
\and U.A.~Y{\i}ld{\i}z\inst{\ref{inst6}}
}

\institute{
Universit\'{e} de Bordeaux, Laboratoire d'Astrophysique de Bordeaux, France; CNRS/INSU, UMR 5804, Floirac, France\label{inst1}
\and
SRON Netherlands Institute for Space Research, PO Box 800, 9700 AV, Groningen, The Netherlands\label{inst10}
\and
Kapteyn Astronomical Institute, University of Groningen, PO Box 800, 9700 AV, Groningen, The Netherlands\label{inst11}
\and           
Ecole des Mines de Nantes, 4 rue Alfred Kastler, 44300 Nantes, France\label{inst46}
\and
Max-Planck-Institut f\"{u}r Radioastronomie, Auf dem H\"{u}gel 69, 53121 Bonn, Germany\label{inst30}
\and
Leiden Observatory, Leiden University, PO Box 9513, 2300 RA Leiden, The Netherlands\label{inst6}
\and
Max Planck Institut f\"{u}r Extraterrestrische Physik, Giessenbachstrasse 1, 85748 Garching, Germany\label{inst2}
\and
Centro de Astrobiolog\'{\i}a. Departamento de Astrof\'{\i}sica. CSIC-INTA. Carretera de Ajalvir, Km 4, Torrej\'{o}n de Ardoz. 28850, Madrid, Spain.\label{inst16}
\and
Observatorio Astron\'{o}mico Nacional (IGN), Calle Alfonso XII,3. 28014, Madrid, Spain\label{inst12}
\and
INAF - Istituto di Fisica dello Spazio Interplanetario, Area di Ricerca di Tor Vergata, via Fosso del Cavaliere 100, 00133 Roma, Italy\label{inst13}
\and
Institute of Astronomy, ETH Zurich, 8093 Zurich, Switzerland\label{inst3}
\and
Department of Astronomy, The University of Michigan, 500 Church Street, Ann Arbor, MI 48109-1042, USA\label{inst14}
\and
Department of Radio and Space Science, Chalmers University of Technology, Onsala Space Observatory, 439 92 Onsala, Sweden\label{inst9}
\and
California Institute of Technology, Division of Geological and Planetary Sciences, MS 150-21, Pasadena, CA 91125, USA\label{inst15}
\and
School of Physics and Astronomy, University of Leeds, Leeds LS2 9JT, UK\label{inst4}
\and
INAF - Osservatorio Astrofisico di Arcetri, Largo E. Fermi 5, 50125 Firenze, Italy\label{inst5}
\and
Astronomical Institute Anton Pannekoek, University of Amsterdam, Kruislaan 403, 1098 SJ Amsterdam, The Netherlands\label{inst17}
\and
Department of Astrophysics/IMAPP, Radboud University Nijmegen, P.O. Box 9010, 6500 GL Nijmegen, The Netherlands\label{inst18}
\and
Department of Physics and Astronomy, Denison University, Granville, OH, 43023, USA\label{inst19}
\and
LERMA and UMR 8112 du CNRS, Observatoire de Paris, 61 Av. de l'Observatoire, 75014 Paris, France\label{inst20}
\and
University of Waterloo, Department of Physics and Astronomy, Waterloo, Ontario, Canada\label{inst21}
\and
Observatorio Astron\'{o}mico Nacional, Apartado 112, 28803 Alcal\'{a} de Henares, Spain\label{inst22}
\and
INAF - Osservatorio Astronomico di Roma, 00040 Monte Porzio catone, Italy\label{inst23}
\and
MPI f$\ddot{u}$r Sonnensystemforschung, D�37191 Katlenburg-Lindau, Germany\label{inst49}
\and
National Research Council Canada, Herzberg Institute of Astrophysics, 5071 West Saanich Road, Victoria, BC V9E 2E7, Canada\label{inst7}
\and
Department of Physics and Astronomy, University of Victoria, Victoria, BC V8P 1A1, Canada\label{inst8}
\and
Centre for Star and Planet Formation, Natural History Museum of Denmark, University of Copenhagen,
{\O}ster Voldgade 5-7, DK-1350 Copenhagen K., Denmark\label{inst24}
\and
Department of Astronomy, Stockholm University, AlbaNova, 106 91 Stockholm, Sweden\label{inst25}
\and
California Institute of Technology, Cahill Center for Astronomy and Astrophysics, MS 301-17, Pasadena, CA 91125, USA\label{inst26}
\and
the University of Western Ontario, Department of Physics and Astronomy, London, Ontario, N6A 3K7, Canada\label{inst27}
\and
Harvard-Smithsonian Center for Astrophysics, 60 Garden Street, MS 42, Cambridge, MA 02138, USA\label{inst28}
\and
Department of Physics and Astronomy, Johns Hopkins University, 3400 North Charles Street, Baltimore, MD 21218, USA\label{inst29}
\and
Jet Propulsion Laboratory, California Institute of Technology, Pasadena, CA 91109, USA\label{inst31}
\and
Department of Physics and Astronomy, University of Calgary, Calgary, T2N 1N4, AB, Canada\label{inst32}
\and
Instituto de Radioastronom\'{i}a Milim\'{e}trica (IRAM), Avenida Divina Pastora 7, N\'{u}cleo Central, E-18012 Granada, Spain\label{inst33}
\and
KOSMA, I. Physik. Institut, Universit\"{a}t zu K\"{o}ln, Z\"{u}lpicher Str. 77, D 50937 K\"{o}ln, Germany\label{inst47}
\and
N. Copernicus Astronomical Center, Rabianska 8, 87-100, Torun, Poland\label{inst48}
}

\date{Received May 15, 2010; accepted July 20, 2010}

\abstract{We present Herschel observations of the water molecule in the massive star-forming region \source. The \hoA, \hoC, \hoB, \hoD, \hoE, and \hoF\ lines, covering a frequency range from 552 up to 1669 GHz, have been detected at high spectral resolution with HIFI. The water lines in \source\ show well-defined high-velocity wings that indicate a clear contribution by outflows. Moreover, the systematically blue-shifted absorption in the H$_2$O lines suggests expansion, presumably driven by the outflow. No infall signatures are detected. The \hoD\ and \hoF\ lines show  absorption from the cold material ($T\sim 10$~K) in which the high-mass protostellar envelope is embedded. One-dimensional radiative transfer models are used to estimate water abundances and to further study the kinematics of the region. We show that the emission in the rare isotopologues comes directly from the inner parts of the envelope ($T \gtrsim 100$~K) where water ices in the dust mantles evaporate and the gas-phase abundance increases. The resulting jump in the water abundance (with a constant inner abundance of $10^{-4}$) is needed to reproduce the \hoA\ and \hoC\ spectra in our models. We estimate water abundances of 10$^{-8}$ to 10$^{-9}$ in the outer parts of the envelope ($T\lesssim100$~K). The possibility of two protostellar objects contributing to the emission is discussed.}

\authorrunning{L. Chavarr\'{\i}a et al.}
\maketitle
%
\section{Introduction}\label{section_introduction}
The water molecule is a key species for studying star formation. In the cool regions of molecular clouds ($T <$ 100~K), water is present as ice in the mantles of dust grains. In the immediate surroundings of high-mass protostars, the dust is heated to temperatures well above 100 K, evaporating the water ices and increasing its abundance in the gas phase by several orders of magnitude and making water one of the most abundant molecules. Understanding how the accretion of matter overcomes radiative pressure in massive protostars is a major astrophysical problem \citep{zinnecker2007}, and water may play an active role in the energy balance \citep[e.g.][]{doty1997}. Due to the abundance jump, water makes it possible to specifically study the inner regions from which the massive protostar accretes.

Low angular-resolution water observations of protostellar envelopes (region $\lesssim 0.1$~pc in size with a density n(H$_2$) $ \gtrsim 10^6$~cm$^{-3}$ and centrally peaked density and temperature profiles) with the Infrared Space Observatory (ISO) and the Submillimeter Wave Astronomy Satellite (SWAS) have revealed overall abundances of less than 10$^{-7}$ \citep{boonman2003}. The much smaller beam of the Herschel Space Observatory \citep{pilbratt2010}, together with the high spectral resolution of the Heterodyne Instrument for the Far-Infrared \citep[HIFI,][]{degraauw2010}, gives a unique opportunity to separate the emission and absorption in the inner ($T\gtrsim 100$~K) and outer ($T\lesssim 100$~K) protostellar envelope in order to build a model of the structure and kinematics of these innermost regions which are critical to fueling the future star.

\begin{figure}
\centering
\includegraphics[width=8cm]{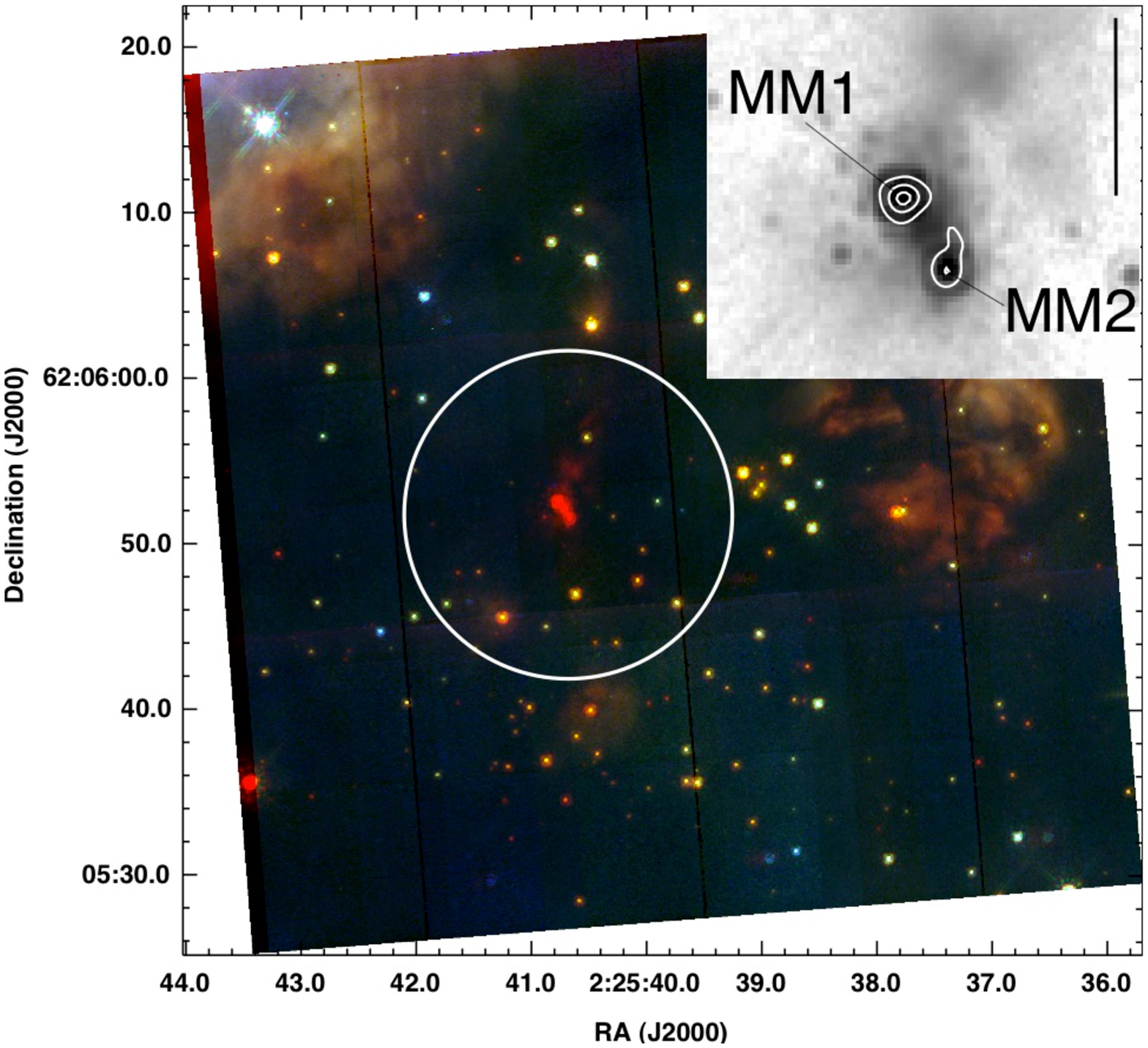}
\caption{NICMOS color-composite image of \source\ and its surroundings from \citet{megeath2005}. The white circle represents the HIFI beam size for the \hoD\ and \hoC\ lines (20$''$). The gray-scale inset is a close-up of the central area at 2.2 microns with white contours corresponding to 30, 60 and 90\% of the peak flux at 1.4 mm \citep{rodon2008}. The black bar (top right of inset) represents a distance of 5.000 AU (2.5$''$).\label{nicmos}}
\end{figure}

In this letter, we report water observations of the region \source\ \citep{wynn-williams1972}. \source\ is a bright infra-red source located in the active star-forming region W3 in the Perseus arm at a distance of 2.0~kpc \citep{hachisuka2006}. Its high FIR luminosity \citep[$1 - 2 \times 10^5$~\lsun,][]{ladd1993} and the detection of radio emission \citep{wilson2003, vandertak2005} indicates that \source\ hosts high-mass stars in an early stage of evolution. High-resolution near-IR images show a cluster of IR-sources in \source\ with two sources identified as the main elements \citep[Figure \ref{nicmos} and][]{megeath2005}. These correspond to the bright millimeter sources MM1 and MM2 identified by \citet{rodon2008}. They are separated by $\sim 1$" and drive two of the several outflows identified by \citet{boonman2003} and \citet{rodon2008}.

The first observations of the \hoB, \hoE\ and \hoF\ water lines and the rare \hoA\ and \hoC\  isotopologues towards a massive protostar are presented here. 

\section{Observations and data reduction}

\source\ was observed with HIFI as part of the Science Demonstration Phase, between March 3-22, 2010. The position observed is at R.A. 02:25:40.6 and Dec +62:05:51 (J2000), close to the peak radio and mid-IR emission \citep{vandertak2005}. The observations are part of the Priority Science Program (PSP) of the Guaranteed-Time Key Program Water In Star-forming regions with Herschel (WISH; van Dishoeck et al, in prep).

Data were taken simultaneously in H and V polarizations using both the acousto-optical Wide-Band Spectrometer (WBS) with 1.1 MHz resolution and the correlator-based High-Resolution Spectrometer (HRS) with 480 kHz resolution (resolutions of 0.30 and 0.13 \kms, respectively, for the 1100 GHz lines). We used the double beam switch observing mode with a throw of 3\arcmin. HIFI receivers are double sideband with a sideband ratio close to unity.

The frequencies, energy of the upper levels, system temperatures, integration times, rms noise level at a given spectral resolution, the beam size and efficiency $B_{\textrm{eff}}$, and the observed continuum level for each of the lines are provided in Table \ref{table_transitions}. Typical uncertainty in the frequencies is on the order of 100~kHz \citep{pearson1991}. Calibration of the raw data onto $T_{\textrm{\tiny{A}}}$ scale was performed by the in-orbit system \citep{roelfsema2010}; conversion to $T_{\textrm{\tiny{mb}}}$ was done with a beam efficiency estimated by raster maps of Mars (Table 1, R. Moreno, priv. comm.). The forward efficiency is 0.96. Currently, the flux scale accuracy is estimated to be between 5 and 10\%. Data was calibrated in the Herschel Interactive Processing Environment \citep[HIPE,][]{ott2010} version 3.0. Further analysis was done within the CLASS\footnote{http://www.iram.fr/IRAMFR/GILDAS} package. After inspection, data from the two polarizations were averaged.

\begin{table*}
\caption{Lines of H$_2$O observed with Herschel/HIFI in \source.}             
\label{table_transitions}      
\centering                          
\begin{tabular}{lccccccccc}        
\hline\hline                 
Water species & Frequency &  $E_u$  &  $T_{\textrm{sys}}$  & $t_{\rm int}$  &  $\delta \nu$ &  rms  &  Beam  &  $\eta_{\textrm{mb}}$  &cont. \\   
                    &    (GHz)     &       (K)  &  (K)    &           (s)            &          (kHz)     & (K)    & (\arcsec) & & (K)\\
\hline                        
   \hoA    & 552.020960   &    61.0  &       73  &  158 & 250 & 0.06 & 40.0 & 0.72 & 0.35 \\
   \hoB    & 987.926764  &  100.8  &       347 &   386 & 500 & 0.07 & 23.0  & 0.71 & 1.4 \\     
   \hoC    & 1101.698256 &    53.4  &       351 &   601 & 500 & 0.09 & 20.0  & 0.71 &2.15 \\
   \hoD    & 1113.342964 &    53.4  &       351 &   601 & 500 & 0.09 & 20.0  & 0.71 &2.15  \\
   \hoE    & 1661.007637 &  194.1  & 1609 & 1164 & 250 & 0.25 & 13.5  & 0.69  & 5.5 \\
   \hoF    & 1669.904775 &  114.4  & 1609 & 1164 & 250 & 0.25 & 13.5  & 0.69  & 5.5 \\
\hline                                   
\end{tabular}
\end{table*}

\section{Results and analysis}
\begin{figure*}
\centering
\includegraphics[width=\textwidth]{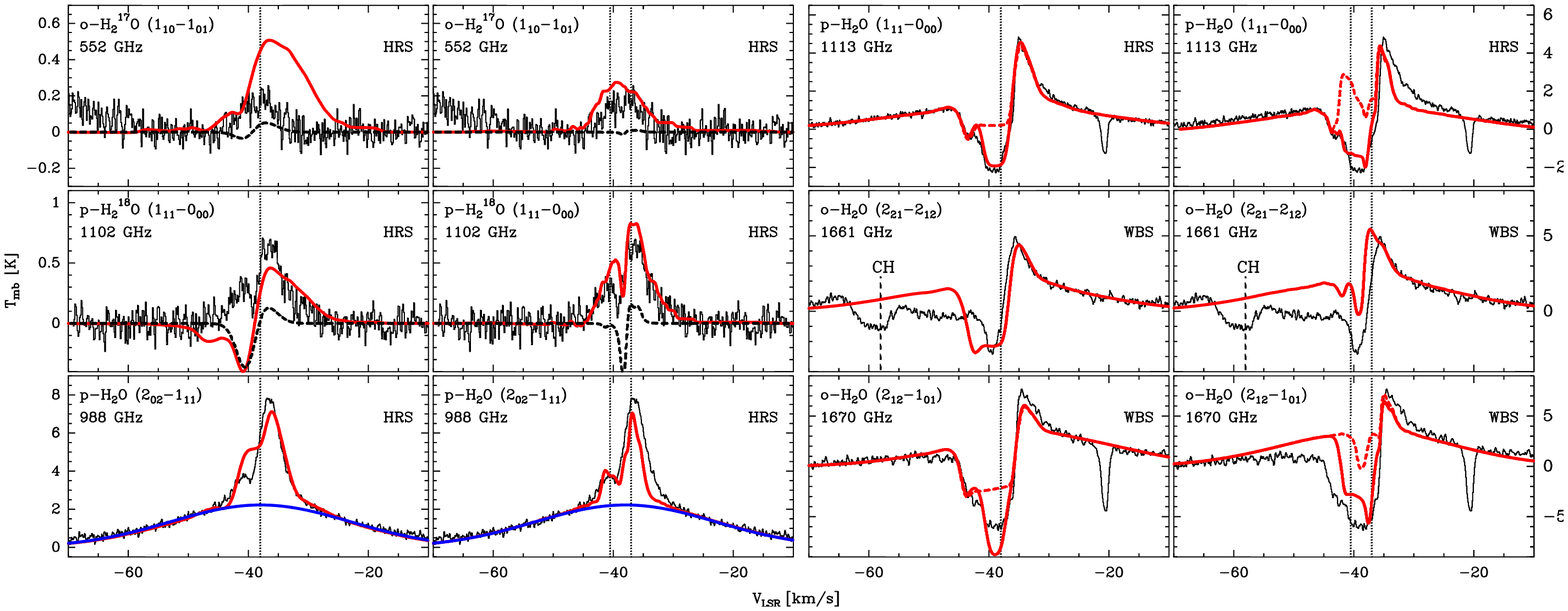}
\caption{HIFI spectra of water lines observed in \source. Columns 1 and 2 show the spectra for the \hoA, \hoC\ and \hoB\ lines, columns 3 and 4 show the spectra for the \hoD, \hoE\ and \hoF\ lines. The single protostellar envelope model fits are shown as red lines over the spectra in columns 1 and 3. The two protostellar envelopes model is shown as red lines over the spectra in columns 2 and 4. The dashed red line over the \hoD\ and \hoF\  spectra show the corresponding model without the cold molecular cloud (see \S~\ref{section_absorption}). The dashed black line over the \hoA\ and \hoC\ spectra shows the model without the abundance jump. Vertical dotted lines indicate the model \vlsr~at --38~\kms\ for the single protostellar envelope model (columns 1 and 3) and --37 and --40.5~\kms\ for the two protostellar envelopes model (columns 2 and 4). Vertical dashed lines in the \hoE\ spectra show the position of the CH line. As an example, the outflow component for the \hoB\ line is shown in blue. The temperature scales are shown on the left side of the figure for columns 1 an 2 and on the right side for columns 3 and 4.}\label{models}
\end{figure*}

Our interpretation of the spectra uses the following terms: inner and outer envelope (as defined in \S~\ref{section_introduction}), cold molecular cloud (the region with $T \sim 10$~K in which the protostellar envelope is embedded), and foreground cloud (unrelated molecular material in the line of sight at a different \vlsr).

The continuum subtracted water spectra are shown in Fig.~\ref{models}. For the \hoA, \hoB, \hoC, and \hoD\ lines, we show the HRS spectra.  For the \hoE\ and \hoF\ lines, we show the WBS spectra because the velocity range covered by the HRS for those lines is not sufficient. The \hoE\ line is contaminated by a CH line (1661.113056 GHz) seen in absorption at $\sim-58$~\kms. The \hoD\ and \hoF\ lines show a narrow absorption feature at $-20.4$~\kms. Since \source\  is located very close to the galactic plane (b $\sim 1.2$\ensuremath{^\circ}), this water absorption is probably produced by a cold ($T \sim 10$~K) foreground cloud on the line of sight. Other cases of absorption due to clouds along the line of sight toward similar sources are reported by \citet{marseille2010} and \citet{wyrowski2010}.

\subsection{Emission and absorption features}\label{section_absorption}

Most of the water emission spectra in \source\ can be described as the sum of two velocity components: one broad (FWHM = 33--40~\kms) and one medium (FWHM = 5--10~\kms). The broad component is visible in all but the rare species and is centered approximately on the \vlsr\ of the source (see the broad component for the 988 GHz line in Fig.~\ref{models}). \citet{boonman2003} observed the outflow in \source\ in CO~$J= 7-6$ with the JCMT and found a similar line width to what is in the water broad component. We assume that the broad component is due to the outflow.

The medium component exhibits emission in the \hoB\ and rare isotopologue lines with two peaks at $-37$ and $-41$~\kms\ in the \hoB\ and \hoC\ spectra. A blend of emission and absorption is seen in the \hoD, \hoE, and \hoF~lines. The absorption in the H$_2$O lines is blue-shifted (P-Cygni profile), suggesting an expansion of the envelope, as also detected by \citet{benz2010} in the hydride lines of \source. The expansion is probably powered by the multiple outflows known to exist in \source\ \citep{rodon2008}. The \hoD\ and \hoF\ lines have a double plateau in absorption at $-38$ and $-42$~\kms. The absorption at $-38$~\kms\ is highly saturated. A simulation using RADEX \citep{vandertak2007} for a cold cloud with a kinetic temperature of 10~K, $N_{\textrm{\tiny{H$_2$O}}}= 3\times10^{13}$ cm$^{-2}$, a line FWHM of 2.8 \kms, and $n_{\textrm{\tiny{H$_2$}}}= 1\times10^4$ cm$^{-3}$ gives opacities of 4.0, 2.2, $2\times10^{-9}$ and $3\times10^{-7}$, respectively, for the \hoD, \hoF, \hoE, and \hoB\ species. This agrees with the detection of saturated absorption only in \hoD\ and \hoF, as well as the absorption by a cold foreground cloud at --20.4~\kms\ also detected only in those lines. The optical depth of the other water lines in cold regions is so low that no absorption is visible. The absorption at --38~\kms\ presumably comes from the cold molecular cloud associated with the source.

The two-component profiles seen in emission in several of the water lines seem to be systematic features of protostellar objects, since it has been observed for low-mass Class 0 objects by \citet{kristensen2010} and for intermediate-mass protostars by \citet{johnstone2010}. \citet{kristensen2010} propose that both the broad and medium components are associated with outflow shocks. In contrast to the low-mass protostars, the emission of the rare isotopologues only comes from the medium component and not from the broad outflow.

\subsection{Two protostellar objects seen in the optically thin lines?}\label{section_2sources}

Double peaks in emission at \vlsr~$\approx -36$ and $-40$~\kms\ in the mostly optically thin \hoB\ and \hoC\ line profiles suggest the detection of more than one protostellar envelope. Since the interferometric observations from \citet{rodon2008} show that there are two millimeter sources within the HIFI beam (Fig.~\ref{nicmos}) that dominate the dust continuum emission in this region, it is possible that the two peaks correspond to emission from these two high-mass protostellar objects.

Ground-based observations of C$^{17}$O, CS, SO, and CH$_3$OH molecules \citep{vandertak2003, vandertak2006, helmich1997} show lines that cover the entire velocity range of emission, but they do not help distinguish between one or two protostellar envelopes. The continuum observations by \citet{rodon2008} could not measure the velocities of MM1 and MM2.

Even though the hypothesis of the detection of two protostellar envelopes in the spectra needs to be investigated in more detail (to be presented in a forthcoming paper), we show in \S~\ref{section_modeling} that a two protostellar envelope model tends to fit the observations better than does a single protostellar envelope model.

\begin{table}
\begin{tiny}
\caption{Parameters of single and two protostellar envelope (PE) models.}             
\label{table_parameters}      
\centering                          
\begin{tabular}{l | r|r r }        
\hline\hline                 
 Parameter  &  Single  PE &  PE 1 &  PE 2 \\    
\hline                        
Outer radius (AU)                 & 12000              & 12000                     & 12000\\
Luminosity (\lsun)               &  10$^5$            & $5 \times 10^4$   &  $5 \times 10^4$  \\
Mass (\msun)                      &  250                   &  125                        &  125     \\
\agua                                   &  $2\times10^{-8}$  &  $1.8\times10^{-8}$  &  $8\times10^{-10}$      \\
Post-jump \agua                 &   $1\times10^{-4}$ &  $1\times10^{-4}$ &  $1\times10^{-4}$ \\
\vtur~(\kms)                      & 2.0                            &  0.4                          &  0.5                \\
\vexp~(\kms)                     & 2.0                            &  1.1                          &  1.2      \\
\vlsr ~(\kms)                      & --38                          &  --37.0                       &  --40.5           \\
\hline                                   
\end{tabular}
\end{tiny}
\end{table}

\subsection{A model of \source}\label{section_modeling}
We used the Monte Carlo code {\it MC3D} \citep{wolf2003} and the radiative transfer program RATRAN \citep{hogerheijde2000} to model the water emission in \source\ following the method described in \citet{marseille2008} with a power-law density exponent of --1.2.  The models assume spherical symmetry. A more detailed investigation of the region's morphology using 2D models is in progress and will be published when the entire set of water lines (including maps) is observed.

Our models in \source\ have three components: an outflow, one (or two) protostellar envelopes, and a cold cloud. The outflow contribution to the spectra was modeled as a Gaussian emission line with FWHM between 33 and 40~\kms. The cold cloud was introduced taking the cloud parameters described in \S~\ref{section_absorption} for the saturated absorption. Three input parameters are used to fit the line profiles in the envelope model: water abundance (\agua), turbulent velocity (\vtur), and expansion velocity (\vexp). The width of the line is adjusted by changing \vtur. The line asymmetry is reproduced by the expansion velocity \vexp. The line intensity is fit by a combination of the abundance, \vtur, and \vexp\ parameters.  The same \vtur\ and \vexp\ have been assumed for all lines. We use abundance ratios of 500 for H$_2$O$/$H$_2^{18}$O, 4.13 for H$_2^{18}$O/H$_2^{17}$O \citep[as measured in the W3 region by][]{penzias1981, wilson1994, wouterloot2008}, and 3 for ortho/para-H$_2$O. The models assume a jump in the abundance in the inner envelope at 100 K with a constant inner abundance of 10$^{-4}$. Having the jump occur at the radius predicted by the physical model leads to overestimating the water line intensity in the inner envelope. We thus decreased the size of the emitting region (to $\sim$1500 AU) by putting the abundance jump at a temperature of 150 K. An alternative model leading to similar results keeps the jump at 100~K and decreases the water abundance in the inner envelope by a factor of 10. We assume that the line intensities from the various components can simply be added as in the optically thin limit, and future models will treat this effect in more detail. Table~\ref{table_parameters} gives the parameters used in the models and the model spectra are shown in Fig.~\ref{models} for both a single and a double protostellar envelope model. The outer radius in the table corresponds to the deconvolved radius taken from the 1.3~mm continuum emission \citep{oldham1994}.

The observed lines were fit using the following strategy. First, we modeled the rare isotopologue lines and then the \hoB\ line (including the outflow emission) using the same \agua, \vtur, and \vexp\ values. Once we were able to reproduce the main features of those profiles by minimizing residuals on a grid of values, we applied the same parameters to the rest of the lines, including an outflow component when justified. For the single protostellar envelope model, it was not possible to correctly fit the rare isotopologues and \hoB\ lines using the same \agua, \vtur, and \vexp\ values. For the double protostellar envelope model, all of the spectra are better fit with the exception of \hoE. The deep absorption in the \hoD\ and \hoF\ lines is reproduced well only if we add a cold cloud to the model. The uncertainties for the water abundances are about a factor of 3 and for \vtur\ and \vexp\ about $\sim1$~\kms.

As shown in Fig. 2, the emission in the rare isotopologue lines is only reproduced in our models by including a jump in the water abundance in the inner envelope. The estimated optical depths for the lines of the rare species are low ($\tau \sim 0.1$, estimated using RATRAN). This suggests that the emission in the rare species comes from the inner envelope where the water abundance is greatly enhanced. Emission from this region could be the expected contribution from the passively (radiatively) heated inner envelope.

\section{Summary} 
Spectra of six water lines observed with the Herschel-HIFI instrument in the high-mass star-forming region \source\ are presented and discussed. A 1D radiative transfer model was applied to estimate water abundances and study the kinematics in the protostellar envelope. The main results follow.
\begin{enumerate}
\item We detect strong water emission and absorption. 
\item A strong outflow component is detected in the water lines. 
\item Water emission shows absorption from cool molecular gas hosting the protostellar envelope. Blueshifted absorption suggests an expansion of the outer envelope, and no infall signature is detected.
\item Radiative transfer models indicate water abundances ranging from 10$^{-8}$ to 10$^{-10}$ in the outer envelope.
\item Based on our model, a jump in water abundance in the inner envelope is needed to reproduce the H$_2^{17}$O and H$_2^{18}$O lines.
\item The optically thin line profiles are better fit using a model with two protostellar envelopes, in agreement with previous interferometric continuum observations of \source.
\end{enumerate}
\begin{acknowledgements}
HIFI has been designed and built by a consortium of
institutes and university departments from across Europe, Canada, and the
United States under the leadership of SRON Netherlands Institute for Space
Research, Groningen, The Netherlands, and with major contributions from
Germany, France, and the US. Consortium members are: Canada: CSA,
U.Waterloo; France: CESR, LAB, LERMA, IRAM; Germany: KOSMA,
MPIfR, MPS; Ireland, NUI Maynooth; Italy: ASI, IFSI-INAF, Osservatorio
Astrofisico di Arcetri- INAF; Netherlands: SRON, TUD; Poland: CAMK, CBK;
Spain: Observatorio Astron\'omico Nacional (IGN), Centro de Astrobiolog\'{\i}a
(CSIC-INTA). Sweden: Chalmers University of Technology - MC2, RSS \&
GARD; Onsala Space Observatory; Swedish National Space Board, Stockholm
University - Stockholm Observatory; Switzerland: ETH Zurich, FHNW; USA:
Caltech, JPL, NHSC.
HIPE is a joint development by the Herschel Science Ground
Segment Consortium, consisting of ESA, the NASA Herschel Science Center, and the HIFI, PACS and
SPIRE consortia.
We also thank the French Space Agency CNES for financial support.
 \end{acknowledgements}

\bibliographystyle{aa} 
\bibliography{biblio} 
\end{document}